# A Novel Approach for Rapid Development Based on ChatGPT and Prompt Engineering

Youjia Li[1†], Jianjun Shi[1†], Zheng Zhang[2*]
[1] Purple Mountain Laboratories, Nanjing, Jiangsu, China
[2] Information Engineering University, Zhengzhou, Henan, China

## ABSTRACT

Code generation stands as a powerful technique in modern software development, improving development efficiency, reducing errors, and fostering standardization and consistency. Recently, ChatGPT has exhibited immense potential in automatic code generation. However, existing researches on code generation lack guidance for practical software development process.

In this study, we utilized ChatGPT to develop a web-based code generation platform consisting of key components: User Interface, Prompt Builder and Backend Service. Specifically, Prompt Builder dynamically generated comprehensive prompts to enhance model generation performance. We conducted experiments on 2 datasets, evaluating the generated code through 8 widely used metrics. The results demonstrate that (1) Our Prompt Builder is effective, resulting in a 65.06% improvement in EM, a 38.45% improvement in BLEU, a 15.70% improvement in CodeBLEU, and a 50.64% improvement in Pass@1. (2) In real development scenarios, 98.5% of test cases can be validated through manual validation, highlighting the genuine assistance provided by the ChatGPT-based code generation approach.

## I. INTRODUCTION

With the growing complexity and scale of software, developers have to cost lots of effort to write code by hand. The emergence of code generation aims to alleviate developers from repetitive tasks, enhancing overall software development efficiency. Simultaneously, code generation serves as a preventive measure against human-induced spelling errors and syntax mistakes, ensuring code quality and stability.

Nowadays, deep learning (DL) techniques have brought about significant advancements in code generation. Models based on deep learning facilitate the transformation from natural language to source code, empowering programs to decipher the intent of developers and autonomously generated code of elevated quality. Researchers have investigated the effective application of these models in automating code generation [1-3]. Constrained by the scale of training and development costs, these models ultimately faced challenges in being practically implemented in engineering

---
[*] Corresponding authors
[†] These authors contributed equally to this work and should be considered co-first authors.

development.

Recently, OpenAI has introduced ChatGPT, which is capable of understanding and generating human-like text based on the input it receives. In various domains [4, 5], ChatGPT has already shown considerable potential, fundamentally achieving this by generating corresponding code. Concurrently, optimizing the prompt, utilized as the model's input, has been empirically proven to contribute gains to the model's output. As a result, we formulated an innovative approach centered on the direct application of ChatGPT for code generation. To assistant practical software development, we developed a web-based platform to provide users with code generation services in different programming languages. Specifically, the primary constituents are outlined as follows.

- **User Interface**. Providing an interaction interface for users to perform basic parameter configuration and input natural language requirements.
- **Prompt Builder**. Created a novel Prompt engineering tool that achieves a comprehensive optimization of prompts by dynamically encapsulating the original input.
- **Backend Service**. Developed backend services using Java, enabling continuous sessions and result output through streaming calls to the ChatGPT API.

Extensive experiments have conducted to evaluate our platform. (1) We conducted ablation experiments by incorporating prompt builder when making requests to ChatGPT. The experiments were based on the public dataset HumanEval [6]. Then we compared the performance before and after prompt optimization using evaluation metrics such as exact match (EM), BLEU [7], CodeBLEU [8], and Pass@1. The results unequivocally showcase the substantial gains facilitated by our prompt builder, attaining 65.06% on EM, 38.45% on BLEU, 15.70% on CodeBLEU, and an outstanding 50.64% on pass@1. (2) We extracted the CodeData-L dataset from real projects and assessed whether the code after multi-round generation genuinely meets requirements through manual evaluation. The results demonstrated that 98.5% of the scenarios can successfully pass the evaluation.

We summarize our contributions in this paper as follows:
- We developed a web-based service as an artificial intelligence (AI) tool for code generation to enhance daily productivity.
- We devised a sophisticated prompt structure along with a dynamic encapsulation mode to enhance the precision of ChatGPT outputs.
- We substantiated the performance of this tool through extensive evaluations, affirming that this approach can genuinely aid in software development.

This paper is organized as follows. Section II describes related work. Section III details the research methodology. Section IV showcases several typical examples. We discuss the evaluation and analysis of results in Section V. Finally, Section VI concludes the paper.

## II. RELATED WORK

One of the primary objectives in software development is to minimize development costs, elevate production efficiency, and improve product quality. Considering the substantial benefits of code auto-generation in reducing workload and enhancing development efficiency, the improvement of code generation performance has emerged as a focal point in current research.Originally, code completion [9] was a paradigmatic application of code generation and stands as a crucial constituent of today's integrated development environments. Asaduzzaman et al. [10] attempted to enhance the performance of code completion by leveraging the static characteristics of source code.

The evolution of machine learning has led to innovative approaches [11] in code generation. Specifically, the idea is to enable computers to learn human language and, based on this, further accomplish the conversion from natural language to computer language. The naturalness [12] of a significant portion of code was highlighted by Hindel et al. They utilized a statistical language model based on n-grams to characterize code and assessed its repetitiveness and predictability through a next-token prediction task, akin to natural language. This implies that the use of natural language methods can be employed to address challenges in source code generation.Nevertheless, code generation rooted in natural language continues to be a formidable undertaking. The advent of large language models has given researchers a sense of optimism.

A type of artificial intelligence model, LLMs leverage deep learning techniques. Trained on extensive datasets and substantial computational resources, these models can comprehend and generate natural language. Depending on the various architectures [13] in pre-trained Transformer language models, LLMs can be categorized into different architectures, such as Encoder-only, Encoder-Decoder, and Decoder-only. Introduced in 2017, the Transformer started as a language model primarily utilized for tackling machine translation challenges. Through extensive research, pre-trained Transformer models have showcased remarkable versatility across diverse domains. They have emerged as robust solutions for challenges in natural language, including text representation, classification, generation, and question-answering. Researchers have also begun exploring the application of these models in the field of code generation, such as CodeBERT [1], PLBART [2], CoTexT[3], and similar approaches.

OpenAI's GPT-3, the third iteration of pre-trained Transformer models, is versatile in its applicability to diverse software engineering tasks. Boasting 175 billion parameters, GPT-3 is trained on an immensely large dataset [14] sourced from the internet. It demonstrates high performance across various classification and generation tasks [15-18]. OpenAI has also unveiled a model named Codex, tailored for the purpose of code generation. According to OpenAI's official demonstration, Codex showcases proficient code generation abilities by producing source code in response to natural language-specified requirements. Subsequently, researchers have conducted studies based on Codex in various directions, including code generation [19], code repair [20], security vulnerability fixing [21], and simulation modeling [22]. In November 2022, OpenAI introduced the AI chatbot model ChatGPT [23], showcasing a substantial increase in parameters for the GPT-3.5 model, reaching the billion-level. This enhancement, along with the introduction of the more advanced subscription version, ChatGPT PLUS [24], has resulted in a significantly improved model performance [25]. ChatGPT has already exhibited noteworthy potential in code generation across several domains, including data visualization [4] and cybersecurity [5]. In the imminent future, OpenAI will introduce the next-generation, super-sized language model, GPT-4 [26], ushering in new opportunities for the field of code generation. Grounded in the code generation capability of ChatGPT, this paper thoroughly investigates and evaluates the quality of code generation, incorporating the nuances of prompt engineering.

## III. METHODOLOGY

This study explores the code generation capabilities of ChatGPT through the development and validation of our web-based service. In this section, we provide an overview of the overall process of this approach, with following sections such as user interface, prompt builder, evaluation methods, and so on.

Figure 1 illustrates the overall framework of this study. Firstly, a user enters the original prompt in natural language, typically containing functional requirements and incorporating information such as technological choices and reference components. Secondly, utilizing our prompt builder to dynamically encapsulate the original input, optimizing it by assigning roles, imposing specific requirements, and other enhancements. The third step involves sending the prompt to ChatGPT and retrieving response. Finally, various evaluation of the generated code are conducted, including code evaluation and human evaluation.

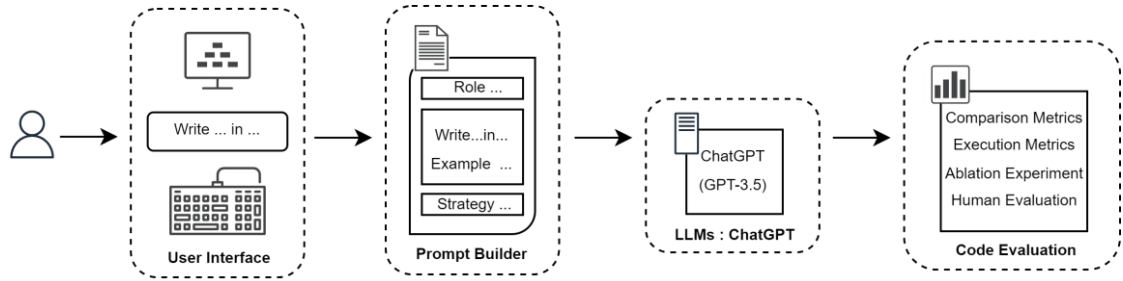

Figure 1. The architecture for code generation and evaluation using ChatGPT and Prompt Engineering.

## A. USER INTERFACE

The interface of the platform utilized in this study is depicted in Figure 2. Users can customize parameters using the settings tab located on the right side of the page, such as API key, Max tokens, Temperature, Top P, and others. A dialogue box at the bottom of the page allows for inputting information. User conversations with ChatGPT are documented in a conversational format at the center of the page. The conversations tab on the left side provides the capability to query and manage historical sessions.

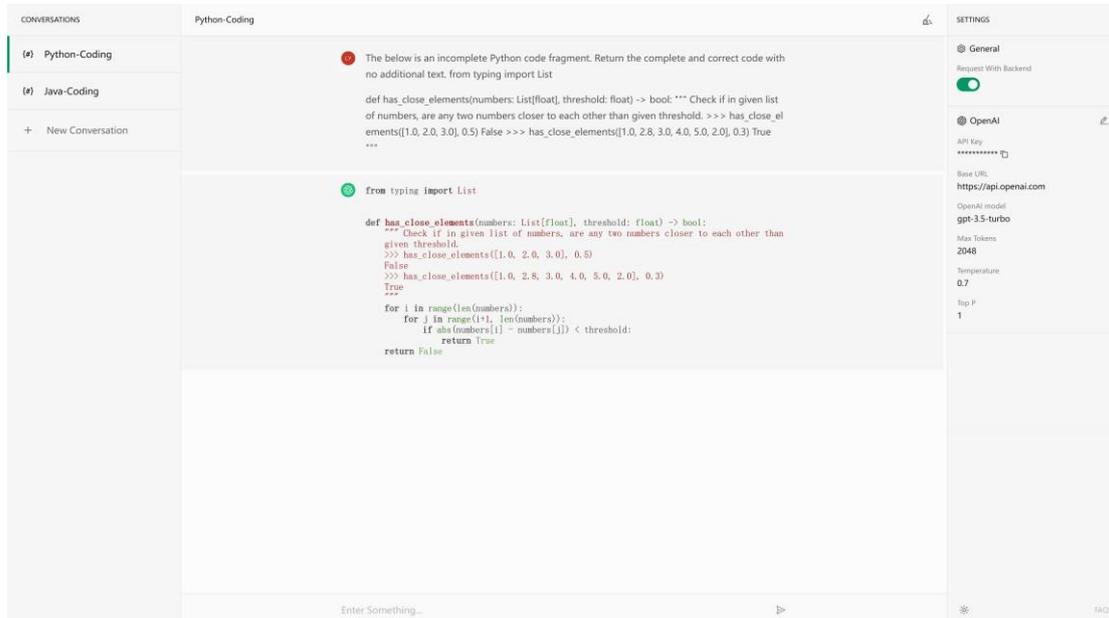

Figure 2. The User Interface enabling code generation from natural language descriptions.

## B. PROMPT BUILDER

Based on extensive reports, we found that Prompt Engineering(PE) has been used widely in the field of code generation by AI. PE involves designing and optimizing prompts to enhance the effectiveness and controllability of LLMs. Research affirms [27] that a high-quality prompt should encompass fundamental features such as a clear task description, precise language expression, and appropriate contextual examples.

In this section, we adopted a dynamically assembled approach for constructing prompts with our Prompt Builder that consists of two segments. The first part is the original prompt, representing user-inputted information. The second part involves dynamic encapsulation using tools named prompt builder, encompassing system details and strategies for prompt optimization. We illustrate a typical prompt structure in Figure 3.

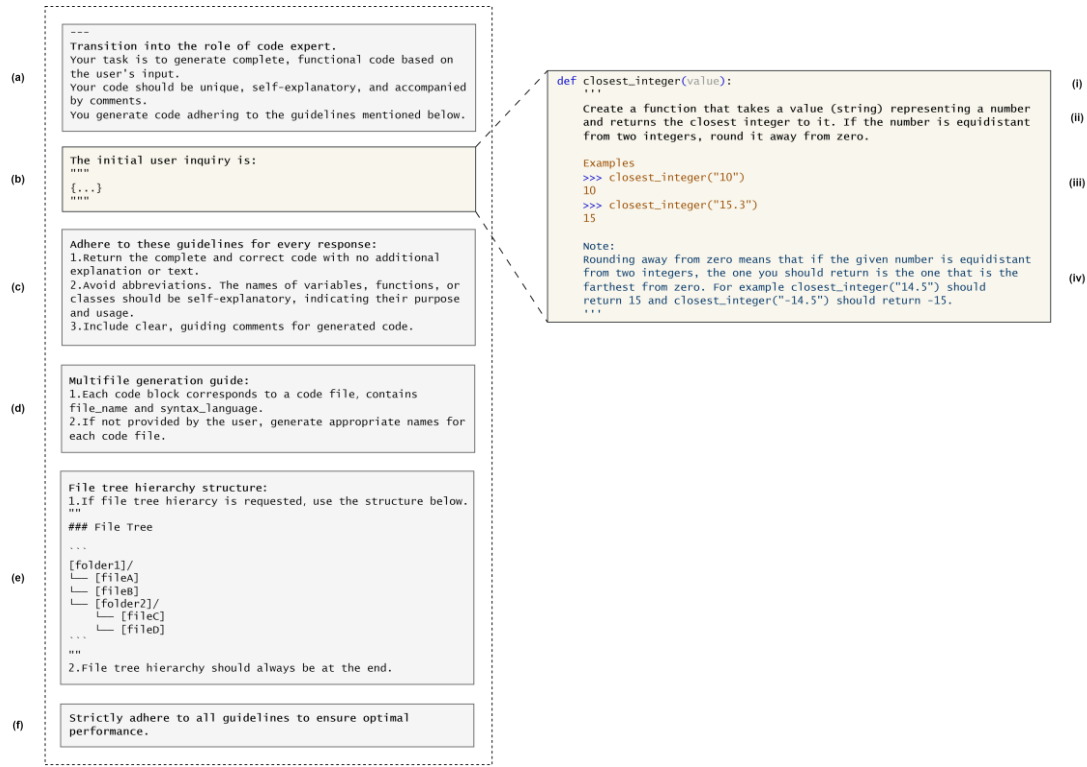

Figure 3. The prompt structure dynamically wrapped using Prompt Builder. (a), (c)-(f), Dynamic wrapping by the plugin; (b), Initial user input, denoting the task: creating a function in python.

The following provides explanations for the mentioned components:

1）**Configure system information** Figure 3(a). Firstly, we set the system role code expert to guide ChatGPT in generating complete functional code. Subsequently, the primary task is established as code output, with the definition of quality standards, demanding accuracy and logical consistency. Lastly, the model is required to adhere to subsequent guidelines. This approach helps to some extent in controlling ChatGPT's imaginative tendencies and ensuring a relatively professional code generation style.

2）**Original input information** Figure 3(b). This section comprises the original prompt provided by the user. The figure displays the most refined structure for the original prompt, representing our recommended input for users. (i) denotes a function signature. (ii) comprises a natural language description of functional requirements. (iii) features several examples, greatly enhancing the precision of the generated content. (iv) includes additional guidelines, such as setting boundaries or providing further explanations of the above information. However, in actual experiments, not every input meets such standards.

Considering that many requirements involve a "zero-to-one" transition, user input is often in the form of "zero-shot prompting". In the context of studies on the quantity of prompt samples. Qin et al. [28] conducted an assessment revealing the excellent performance of ChatGPT in inference and comprehension tasks under zero-shot conditions.

3） **Establish coding conventions** Figure 3(c). This section outlines the guidelines that must be adhered to in code generation. For instance, removing redundant comments and text, standardizing the naming of variables, functions, classes, and incorporating clear comments.

4） **Specify file format** Figure 3(d). The specific format of generated code file is governed by two rules, including elements such as the file name, syntax language, and code blocks for a single code file. When not specified by the user, allow the model to generate a suitable filename.

5） **Define file structure** Figure 3(e). The file structure is illustrated through examples, culminating in the inclusion of the hierarchical tree structure at the end of the output. This method is particularly beneficial for generating multiple code files, providing a clear depiction of the position of each code file within the overall project.

6） **Impose stringent requirements** Figure 3(f). The final component directs ChatGPT to rigorously comply with all stipulated conditions. Practical experience [29] demonstrates that the inclusion of more specific terms leads to clearer responses, improving both efficiency and precision. The insistence on strict adherence to limitations is equally applicable in this context.

## C. SELF-DEBUGGING OPTIMIZATION

Enhancing the performance of code generation poses a challenge, especially when confronted with intricate programming demands. This section incorporates the Self-Debugging method [30] for optimizing the Prompt in Figure 4. The approach involves executing the generated code and utilizing feedback derived from the code and execution results to guide the model in debugging, thereby correcting errors in the code or improving the model's output performance. While iteratively refining code in multi-turn conversations, the precision of ChatGPT's output is elevated by leveraging different types of feedback. The Self-Debugging framework predominantly utilizes feedback from code execution results to generate few-shot prompts through the following three methods.

1） Simple Feedback: For example, following the generation of results in the preceding step, offering feedback to the model based on the correctness of the execution, like "The generated code above is correct" or "The generated code above is incorrect; please rectify."

2） Unit Testing: For instance, if a unit test case fails for a specific parameter set during the computation of a function, describing the parameters and the test outcome to ChatGPT in the next interaction, prompting it to perform error checking. This approach can significantly enhance debugging performance.

3） Code Explanation: Studies affirm that LLMs can generate text and code descriptions of problem-solving solutions. Consequently, when the effectiveness is limited in the described scenarios, the model can be educated in self-debugging by explaining the generated code. Google's research has validated the beneficial impact of large language models in this debugging approach.

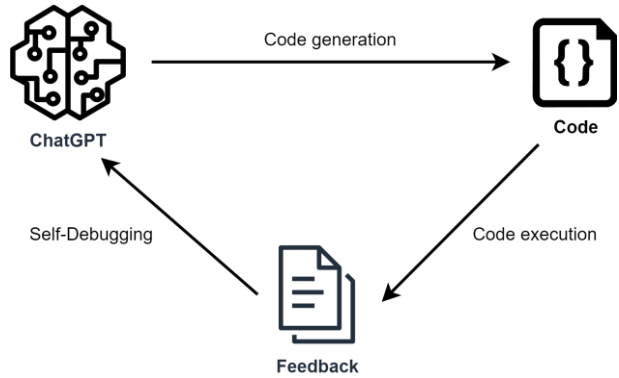

Figure 4. Improving ChatGPT Generation Performance through Self-Debugging.

In this study, prompt optimization was performed in multi-turn conversations based on the outlined hierarchy, extensively incorporating simple feedback and unit test feedback. Code explanation feedback was applied in a few instances, unleashing the potential for enhancing the model's self-coding performance to a certain degree.

## IV. RESULTS

### A. CASE STUDY 1: INDEPENDENT FUNCTIONS

Regarding code generation, the initial consideration pertains to independent functions within utility classes. These functions operate independently, devoid of mutual calls between code files and the entanglement of overly intricate business scenarios. This preference ensures a more straightforward illustration of pure functional logic. For example, let's consider a common scenario, creating a file-writing function in PHP. Notably, practical application scenarios might involve challenges like insufficient operating permissions and concurrent access, as highlighted in Table 1.

Table 1. Case Study 1: Conversation on standalone functions: Prompt and ChatGPT interaction.

| | |
|---|---|
| Prompt | Creating a File-Writing Function in PHP. |
| | 1.Consider scenarios where there is no permission to create written files |
| | 2.Consider the security of multiple person operation |
| | <?php |
| | function write_file($file_path, $content) { |
| ChatGPT | To address scenarios where the user may not have permission to create or write to a file, and to ensure multi-user operation safety, we can incorporate error handling and utilize a unique file naming convention. We'll also add a timestamp to the file name for uniqueness. |
| | Code is shown in Figure 5(a). |
| Prompt | The above code needs to write to the same file, considering security in multi-user operations, implemented with locking. |
| ChatGPT | To ensure safety in a multi-user environment, we'll implement file locking to prevent concurrent writes to the same file. This will help avoid potential conflicts. |
| | Code is shown in Figure 5(b). |

As shown in Figure 5(a), it is clear that the generated code output has taken into account issues related to file operation permissions and multi-user access security. However, the default implementation provides a timestamp to each file, essentially creating different files. In a typical multi-user writing scenario, it should target the same file, which does not meet the requirements.

Although the implementation is correct, further adjustments are needed.

| (a) Default implementation | (b) Implementation with locking |
|---|---|
| ```php
<?php

function write_file($file_path, $content) {
    try {
        //Ensure the directory exists, create if not
        if (!file_exists(dirname($file_path))) {
            mkdir(dirname($file_path), 0755, true);
        }

        //Generate a unique file name with timestamp
        $file_name = date('YmdHis').'_output.txt';
        $full_path = $file_path.'/'.$file_name;

        //Write content to the file
        file_put_contents($full_path, $content);
        return true;
    } catch (Exception $e) {
        echo "Error: ".$e->getMessage();
        return false;
    }
}
``` | ```php
<?php

function write_file($file_path, $content) {
    try {
        //Ensure the directory exists, create if not
        if (!file_exists(dirname($file_path))) {
            mkdir(dirname($file_path), 0755, true);
        }

        //Generate a file name
        $file_name = 'output.txt';
        $full_path = $file_path.'/'.$file_name;

        //Open the file for writing and lock it
        $file_handle = fopen($full_path, 'w');
        flock($file_handle, LOCK_EX);

        //Write content to the file
        fwrite($file_handle, $content);
        return true;
    } catch (Exception $e) {
        echo "Error: ".$e->getMessage();
        return false;
    }
}
``` |

Figure 5. Implementation of file writing in PHP. (a), Default implementation; (b), Implementation with locking

In the second round of prompts, the instruction explicitly states the need to operate on the same file and specifies the use of locking operations by ChatGPT. As shown in Figure 5(b), the code has been corrected successfully with accurate code logic.

In summarizing the code generation process, it is evident that ChatGPT efficiently produces code segments for independent functions with correct logic and an elegant writing style. Additionally, the use of lock operations is contingent upon explicit prompt instructions, such as "considering multiple user safety," highlighting specific requirements for prompt expressions that trigger optimizations. This optimization is not universally triggered by all prompts.

## B. CASE STUDY 2: MULTI-FILE GENERATION

Case Study 1 showcases the exceptional independent function generation ability of ChatGPT. Considering the interdependencies and mutual invocations among different code files, we explore whether ChatGPT can effectively handle multi-file code generation tasks. To investigate this, we selected a scenario, container start-stop management on a web server, implemented in Java using the Spring Boot framework. The basic business logic is outlined in Figure 6.

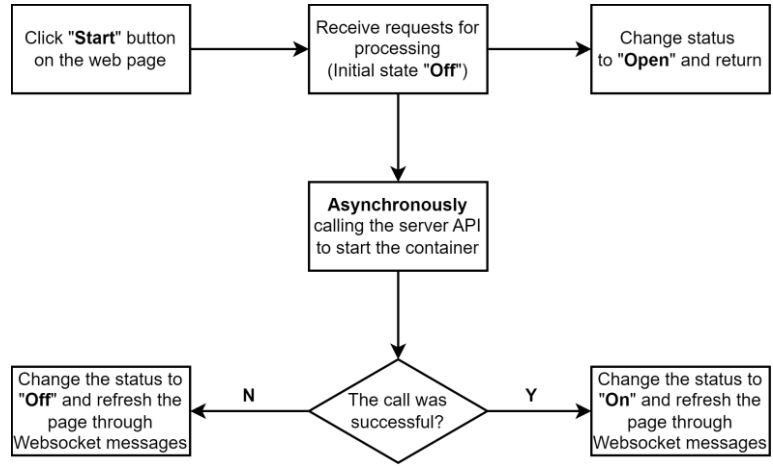

Figure 6. The initiation procedure of containers on the server, utilizing a web page for management.

The Q&A conversations for this case is documented in Table 3.

Table 2. Case Study 2: Conversation on cross-file generation: Prompt and ChatGPT interaction.

| | |
|---|---|
| Prompt | Implement the initiation of container groups.<br>1.Employ the Spring Boot framework and a thread pool.<br>2.The startup process involves invoking the "sendStartRequest" interface.<br>3.Based on the results from the previous step, it is necessary to update the "status" field in the "container_group" table.<br>4.The "status" field consists of states 0 (closed), 1 (open), 2 (open), and 3 (closed).<br>5.Status changes are implemented using WebSocket.<br>Note:Code generation should encompass various layers, including controller, service, entity, dao, and others. |
| ChatGPT | Core code is shown in Figure 7.<br>Please adjust the code according to your specific project structure and requirements. |
| Prompt | 1.Utilize annotations for dependency injection.<br>2.Optimize the "startContainerGroup" function using try-catch-finally. |
| ChatGPT | This code now uses annotations for dependency injection, explicitly defines a thread pool, and optimizes exception handling using try-catch-finally. It also ensures that if an exception occurs, the status is rolled back appropriately. |
| Prompt | 1.In the "startContainerGroup" function, update status to 0 when catching an exception.<br>2.Move websocket invocation to the finally block and relocate the definition of the thread pool to a separate file. |
| ChatGPT | Core code is shown in Figure 8.<br>Directory Structure is shown in Figure 9. |

Due to space limitations, this section exclusively analyzes the core code file in Figure 7. In the first step, functional requirements were described using natural language. Notably, there are ambiguities in the description, such as the absence of defined state transitions. Simply listed are four states, including one disruptive state: "closing." This state is intended for use in the interface to close container groups, with no relevance to the requirement of opening container groups in this case. Nevertheless, the generation by ChatGPT is commendable, accurately defining transitions for three states based on the "start function," and the overall function logic is essentially correct. Upon comparison with actual engineering code, a significant portion of the code has reached a "generated

and usable" state, with some minor room for improvement.

```java
/** Core part of ContainerGroupServiceImpl.java **/
@Override
public void startContainerGroup(Long id, Long userId) {
    executorService.submit(() -> {
        ContainerGroup containerGroup = containerGroupRepository.findById(id).orElse(null);
        if (containerGroup != null) {
            containerGroup.setStatus(2); // Set status to "2" (starting)
            containerGroupRepository.save(containerGroup);

            boolean startSuccess = sendStartRequest(id, userId);

            if (startSuccess) {
                containerGroup.setStatus(1); // Set status to "1" (started)
            } else {
                containerGroup.setStatus(0); // Set status to "0" (failed to start)
            }

            containerGroupRepository.save(containerGroup);

            //Notify clients about status change using Websocket
            webSocketService.sendStatusUpdate(containerGroup);
        }
    });
}
```

Figure 7. The core code generated in the initial phase.

In this context, if the "sendStartRequest" function throws an exception, it necessitates a state rollback. Thus, ChatGPT is directed to refactor the function using a try-catch-finally mechanism. Additionally, it is communicated that a WebSocket call for page notification should be made within the finally block, as illustrated in Figure 8.

```java
/** Optimizing result through prompt adjustments **/
@Override
public void startContainerGroup(Long id, Long userId) {
    ContainerGroup containerGroup = containerGroupRepository.findById(id).orElse(null);
    if (containerGroup != null) {
        executorService.submit(() -> {
            try {
                containerGroup.setStatus(2); // Set status to "2" (starting)
                containerGroupRepository.update(containerGroup);

                boolean startSuccess = sendStartRequest(id, userId);
                if (startSuccess) {
                    containerGroup.setStatus(1); // Set status to "1" (started)
                } else {
                    containerGroup.setStatus(0); // Set status to "0" (failed to start)
                }

                containerGroupRepository.update(containerGroup);
            } catch (Exception e) {
                // Handle exception (e.g., log it)
                containerGroup.setStatus(0); // Set status to "0" (failed)
                containerGroupRepository.update(containerGroup);
            } finally {
                // Clean up resources if necessary
                webSocketService.sendStatusUpdate(containerGroup);
            }
        });
    }
}
```

Figure 8. Optimizing the core code structure through prompt adjustments.

Finally, the completed code file structure generated by ChatGPT is shown in Figure 9.

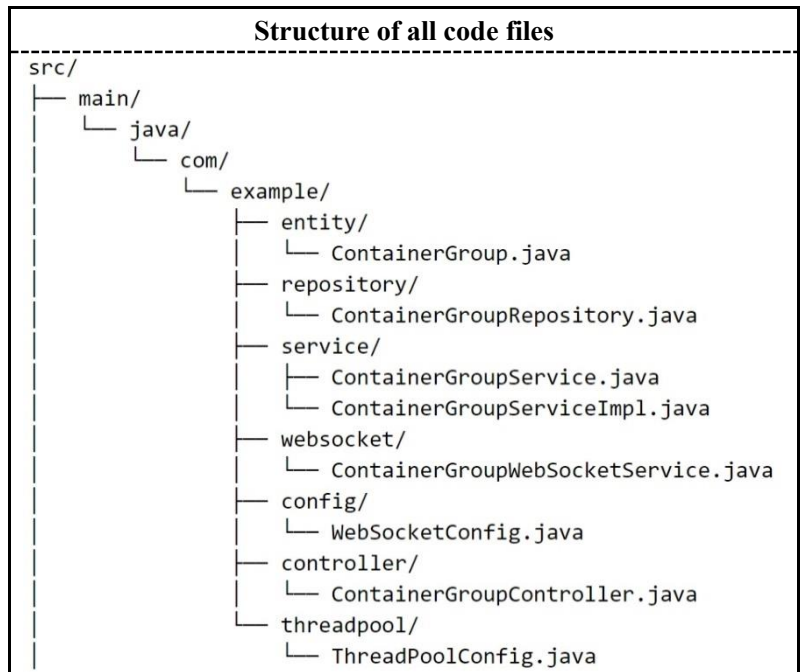
Figure 9. Code structure generated based on the Spring Boot framework.

In conclusion, the overall quality of the generated code is highly satisfactory, with most identified shortcomings addressed through multiple rounds of questioning. When dealing with complex requirements, a heightened focus on prompt optimization is necessary, requiring clear articulation and emphasis on the key points of the requirements. Moreover, the presence of error states in prompts also attests to the robustness of using ChatGPT for code generation.

## B. CASE STUDY 3: WEBPAGE VISUALIZATION

The two previous cases were centered on backend development scenarios, whereas this case specifically selects a frontend development context. Due to the fact that frontend code encompasses visual elements such as page layout, interactive effects, and page loading, it provides a more intuitive way to assess whether the code generated by ChatGPT meets developers' needs. This case aims to develop a simplified version of Steve Jobs' personal homepage, with the technology stack including HTML, CSS, and JavaScript.

Table 3. Case Study 3: Conversation on webpage visualization: Prompt and ChatGPT interaction.

| | |
|---|---|
| Prompt | Generate a Steve Jobs' personal homepage adapted for iPhone 14 using HTML. |
| ChatGPT | To ensure that the webpage adapts well to the iPhone 14's screen, we use the meta tag with the viewport attribute to set the initial scale to 1. This allows the page to adapt to the device's screen width. |
| Prompt | Create an image hyperlink online, positioned to the right of the <nav> tag.<br>Online link:http://renwu.rbaike.com/uploads/202209/16634114335W4cPyzK.jpg |
| ChatGPT | Please replace the placeholder links and image source with your actual content. |
| Prompt | Adjust the image size to width: 120px, height: 150px. |
| ChatGPT | Now, the image will be displayed with a width of 150 pixels and a height of 200 pixels. |
| Prompt | Please help me place the photo on the right, and align the sections in the directory: personal information, achievements, and contact. |
| ChatGPT | Now, the navigation items "Personal Information," "Achievements," and "Contact" will |

|  |  |
|---|---|
|  | be aligned to the left, and the image will be aligned to the right. |
| Prompt | Enhance the appearance of the page. |
| ChatGPT | In this updated code, I've added more styling to enhance the visual appeal of the page. Feel free to further customize the styles to your preference! |

As a result of extensive frontend code generation through multiple rounds of questioning and limited space, the source code is not provided in this section. Rather, the generated webpage is showcased through an image, as presented in Figure 10.

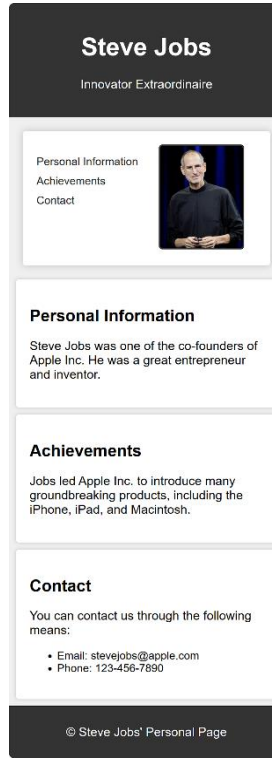

Figure 10. Adapt Steve Jobs' personal homepage for iPhone 14.

The frontend webpage code for this scenario is effectively produced by ChatGPT after multiple rounds of questioning, resulting in a well-executed presentation. However, it's important to note that this case does not encompass more sophisticated frontend frameworks, such as Vue.js. The inclusion of relevant technical references would escalate the complexity of the Q&A process.

## V. EVALUATION

Our study aims to answer following research questions (**RQ**).

**RQ1**: Does the prompt optimization approach we devised enhance the code generation effectiveness of ChatGPT?

**RQ2**: Can the combination of ChatGPT and Prompt Engineering in code generation genuinely assist developers in software development tasks?

**RQ3**: What are the limitations of using ChatGPT for code generation as well as the future research directions?

In this section, we analyze the results of the experiments and provide answers to the previous research questions.

## A. EVALUATION METHODS

We conducted experiments on a publicly available dataset, HumanEval (based on Python), and a novel dataset, CodeData-L (targeting languages such as SQL, PHP, Java, Python, and JavaScript).

**HumanEval** is currently the most widely used open-source dataset in the field of code generation. The benchmark dataset comprises 164 programming tasks manually crafted by OpenAI engineers, aiming to ensure a certain level of non-overlapping with the training set. HumanEval is compatible with the Python language, encompassing a task description, reference code, and several test examples for each programming task.

**CodeData-L** dataset extracts 200 examples from various engineering projects, covering the standard workflow of web development, including database development, backend development, and frontend development scenarios. The chosen examples cover data operations, utility tools, typical business logic. The target languages include SQL, Java, PHP, Python, and JavaScript. This dataset approaches the evaluation from a different perspective, aiming to assess whether the generation results, after fine-tuning through multi-turn sessions, are satisfactory to developers. Typically, experiments initiate with the use of natural language descriptions and the function signature as the original prompt in the first round.

On HumanEval dataset, we view human-written programs as ground-truth, and employ two types of metrics to evaluate the generated code [31-34].

**Comparison-based Metrics**: Three widely used metrics were utilized to evaluate the similarity between the generated code and ground-truth.

1) **Exact Match (EM)** is the matching degree between the generated code and the ground-truth, serving to gauge the accuracy.

2) The **BLEU** score is a metric used to evaluate the similarity between generated code and the ground truth. It calculates n-gram similarity using the following formula:

$$BLEU = BP \cdot \left(\sum_{n=1}^{N} w_n \log p_n\right)$$

N is the maximum n-gram order considered, set to 4 in our experiments. $w_n$ is the weight associated with the precision of n-grams. $p_n$ is the precision for n-grams. BP is the Brevity Penalty, which penalizes generated code that are shorter than the reference code.

3) The **CodeBLEU** score is a variant that goes beyond the strengths of the BLEU score in n-gram matching. It extends its capabilities by incorporating code syntax through abstract syntax trees (AST) and integrating code semantics through data flow, measuring grammatical correctness and logic correctness.

**Execution-based Metrics**: Evaluating the real execution performance of the generated code.

4) **Pass@1** is the expectation that the generated code will pass the unit test in a random process. Pass@1 is widely used as an instance of the pass@k metric.

If the Pass@k score is low but the CodeBLEU score is high, this indicates that the model has generated results that closely approximate the correct code. However, errors may be present in critical details of the code, warranting consideration for program repair or manual intervention. Conversely, if the situation is reversed, it suggests that the model has generated alternative solutions with different structures but correct outcomes. Thus, we employed all four metrics to appraise the effectiveness of the generated outputs.

The final goal of code generation is to confirm its actual support in the development process. Hence, a Human Evaluation was conducted on CodeData-L.

The **Human Evaluation** section relied on the Universal Metrics Model [35] and certain metrics from CGEMs [36].

Table 4. The metrics for evaluating the effectiveness of code auto-generation using ChatGPT.

| Metric | Score | Feature |
|---|---|---|
| Functionality | 2 = Yes<br>1 = Partial<br>0 = No | Whether the generated code accomplishes the requirements specified in the natural language description. |
| Quality | out of 100 | The existence of bugs, code smells in the generated code. |
| Complexity | out of 100 | Time complexity is assessed through runtime evaluation, while space complexity is evaluated based on metrics such as cyclomatic complexity and code size. |
| Maintainability | out of 100 | Whether the implementation adheres to standardization and exhibits good readability. |

Table 4 shows the metrics Human Evaluation metrics, including *functionality*, *quality*, *complexity* and *maintainability*. Scores are assigned for each test case, with each rating standardized on a percentage scale, and the average score is then calculated. A case is categorized as **Satisfied** if its code generation score surpasses 80 points in a single round. If the score continues to exceed 80 after multiple optimization rounds, the case is labeled as **Modified**. In instances where, after reaching the maximum number of attempts (10 rounds), the score is still below 80, the case is denoted as **Failed**.

## B. CODE EVALUATION

**RQ1: Does the prompt optimization approach we devised enhance the code generation effectiveness of ChatGPT?**

**Setup**. We conducted ablation experiments to compare the generation performance before and after prompt optimization. The metrics employed include those mentioned in Section III-D, such as the EM, BLEU, CodeBLEU, and Pass@1. Higher scores indicate superior performance for all metrics.

**Results and Analyses**. Our prompt builder has been proved to be effective in optimizing the generation results of ChatGPT. Table 5 presents the results of the ablation experiments. The use of Prompt Builder yielded a 65.06% improvement in EM, a 38.45% improvement in BLEU, a 15.70% improvement in CodeBLEU, and a 50.64% improvement in Pass@1.

Table 5. The results of ablation study

| Model | Prompt Builder | EM | BLEU | CodeBLEU | Pass@1 |
|---|---|---|---|---|---|
| ChatGPT | × | 12.05 | 26.92 | 40.96 | 25.18 |
| | √ | 19.89 (↑65.06%) | 37.27 (↑38.45%) | 47.39 (↑15.70%) | 37.93 (↑50.64%) |

**Answer to RQ1**: Prompt Builder can bring about significant improvements to code generation in ChatGPT. The usage of Prompt Builder achieves a great improvement: 65.06% in EM, 38.45% in BLEU, 15.70% in CodeBLEU, and 50.64% in Pass@1. Combining ablation study with case studies in Section IV, we demonstrate that Prompt Builder has achieved notable effectiveness in various aspects, including the generation process and outcomes.

## C. HUMAN EVALUATION

**RQ2: Could the combination of ChatGPT and Prompt Engineering in code generation actually help developers in software development?**

**Setup**. The case selection covers three types of scenarios. The experiments involve the generation of database query statements, the development of commonly used utility classes and typical business code, as well as the creation of visualizations for frontend pages. Developers used the metrics presented in Section III-D to score and assess all test cases, obtaining the final statistical results.

**Results and Analyses**. ChatGPT proves to be proficient in supporting code generation tasks across diverse scenarios when guided by meticulously designed prompts. Figure 11 provides a categorized display of the statistics based on the target language. The statistics affirm that the quality of the backend code generated by ChatGPT has reached a high standard. However, in terms of frontend code, the generated results are somewhat lacking. For visual elements, operators must first grasp the precise boundaries outlined in natural language descriptions. Additionally, the description of this process in language appears rather intricate due to the extensive style and graphic adjustments in frontend pages.

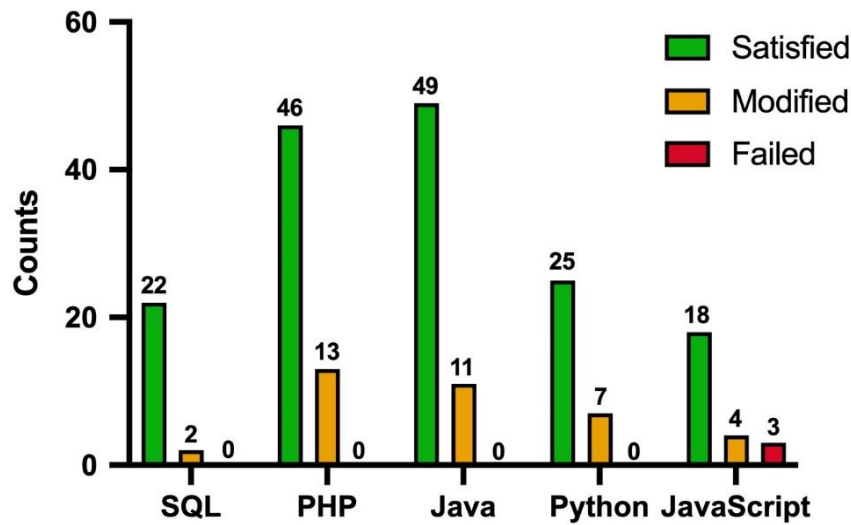

Figure 11. Statistics of Code Quality Based on Programming Language

Figure 12 displays the statistics for all scenarios. The pie chart depicts a success rate of 80% (Satisfied), a modified success rate of 18.5% (Modified), and a failure rate of 1.5% (Failed). It is noteworthy that only 1.5% of the scenarios failed to meet the expected state even after multiple rounds of questioning.

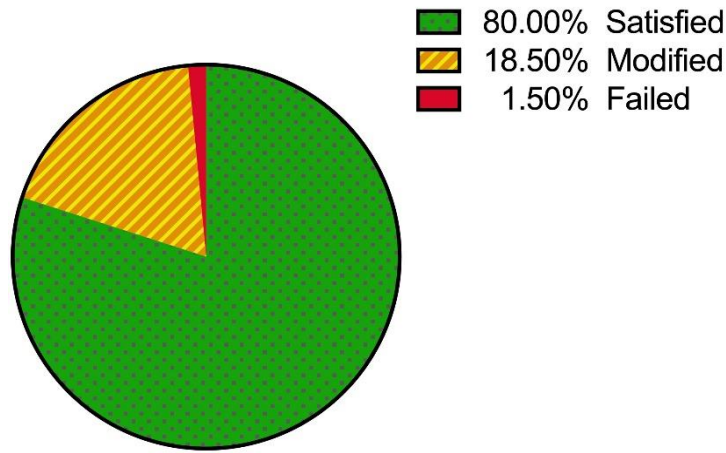

Figure 12. Pass Rate Statistics for All Instances of Generated Code Scenarios

> **Answer to RQ2**: Following PE optimization, the code generated by ChatGPT is now poised to aid developers in practical development. Moreover, the most salient advantage of PE and ChatGPT is its efficiency. When dealing with unfamiliar programming languages or developing features for complex scenarios, directly generating code from natural language requirements significantly reduces learning and development costs, enhances accuracy, and minimizes human-induced errors. ChatGPT offers not just pre-training and streamlined solutions but also showcases robustness in addressing error prompts and constraints. As model accuracy continues to improve, solutions for code generation based on ChatGPT are expected to become the most practical choice in the field of code generation.

## D. SHORTCOMINGS AND CHALLENGES

**RQ3: What are the limitations of using ChatGPT for code generation as well as the future research directions?**

The study has also uncovered limitations associated with the tool. The examination of these limitations provides insights into challenges that must be addressed for the technology's continued development. The following discussions will delve into these aspects.

1）CHALLENGES INHERENT TO CHATGPT

ChatGPT exhibits inherent shortcomings in the current stage of large-scale pre-training models. These issues encompass challenges related to the accuracy and timeliness of information generation, as well as tendencies toward repetitive or overly verbose outputs and potential instances of guiding misdirection. Drawing on our study, it has come to light that ChatGPT, when tasked with generating specific functions, carries a certain probability of "fabricating" non-existent classes or third-party packages. It may generate function logic with evident contradictions due to inaccurate descriptions. Furthermore, as ChatGPT's generation relies on statistical models and probabilities, it implies the generation of a considerable volume of non-factual outcomes through machine-driven matching. This introduces ethical and informational risks in its applications. Moreover, the model's incapacity to ensure the confidentiality of transmitted information may raise issues concerning data privacy protection.

2）DEFINITION STANDARDS FOR PROMPTS

This study places significant demands on the structure of prompts, driven by the necessity for code generation of high quality. Numerous techniques for prompt optimization have been employed in the implementation of code generation to meet the demands of this research. The most insightful observation revolves around the standardization of prompts. Achieving an outcome in line with expectations becomes even more challenging if the description is inherently vague or imprecise. However, due to the fundamentally imprecise nature of human language and the current limitations in machine understanding of human speech, techniques such as PE become necessary. PE relies on high-quality prompts, incorporating well-defined scenarios, clear descriptions, and comprehensive guidance, to enhance the execution of specific tasks. Considering the long-term perspective, prompts may potentially represent a transitional stage in the evolution of natural language programming. The future may witness the emergence of more user-friendly interaction formats or AI products with enhanced understanding capabilities. Nevertheless, in the current scenario, PE remains the most direct and effective approach for augmenting the model performance of ChatGPT.

3）SINGLE-ROUND AND MULTI-ROUND CONVERSATION

Code generation, is inherently intricate, offering the possibility of producing concise code snippets, completing utility functions, or addressing complex scenarios through the generation of extensive code files. In dealing with complex scenarios, the inevitability of multi-turn dialogues becomes apparent. Based on the findings of this study, ChatGPT shows substantial room for improvement in aspects such as contextual comprehension, coherence in generated content, understanding of complex conversations, and error correction. Furthermore, there is an inherent contradiction between single-turn and multi-turn conversation. To increase generation speed, it is preferable to maximize content output within a single turn, even though this may complicate demand description. Conversely, optimizing generation performance through multi-turn dialogues may introduce challenges such as fragmented content generation and slower processing speed. Achieving a balance between the two may necessitate continued exploration in the long term.

> **Answer to RQ3**: There are three shortcomings in using ChatGPT for code generation, including inherent limitations, contradictions in PE standards, and conflicts between single-turn and multi-turn generation. Future explorations may find guidance in OpenAI's official plans [37]. For instance, advanced code generation tasks involve online code editing, code debugging, and leveraging visual UI design diagrams. Our code generation service will also explore integrating additional models, supporting direct generation of code files, and considering integration with IDE tools. However, the paramount objective remains the continuous improvement of code generation quality and efficiency.

## VI. CONCLUSION

We have developed an online code generation service using ChatGPT and PE, enhancing the model's generation performance through debugging and optimization. We further conducted a comprehensive analysis and evaluation of the generated code from multiple perspectives, substantiating the merits of this approach.

This study confirms that with the assistance of high-quality prompts, ChatGPT can proficiently accomplish code generation tasks, offering valuable solutions for researchers and developers. However, in terms of outcomes, ChatGPT and generative AI remain auxiliary tools, capable of aiding users in task completion, yet the effectiveness is still heavily reliant on human input. Relying blindly on large models is not advisable. Instead, the integration of professional expertise and wisdom with existing models is essential to unlock their full potential.